# Uploading Brain into Computer: Whom to Upload First?

| Yana B. Feygin | Kelly Morris | Roman V. Yampolskiy |
| --- | --- | --- |
| Peds-CHSRU | CECS | CECS |
| University of Louisville | University of Louisville | University of Louisville |
| yana.feygin@louisville.edu | kelly.morris.2@louisville.edu | roman.yampolskiy@louisville.edu |

**Abstract**
The final goal of the intelligence augmentation process is a complete merger of biological brains and computers allowing for integration and mutual enhancement between computer's speed and memory and human's intelligence. This process, known as uploading, analyzes human brain in detail sufficient to understand its working patterns and makes it possible to simulate said brain on a computer. As it is likely that such simulations would quickly evolve or be modified to achieve superintelligence it is very important to make sure that the first brain chosen for such a procedure is a suitable one. In this paper, we attempt to answer the question: Whom to upload first?

## 1. Introduction

As we write this paper there is a team of researchers who are working toward the creation of a "Brain Simulation Platform", software that will map the human brain down to a minute level of detail (see https://www.humanbrainproject.eu). This research has incredible implications for many scientific fields of study. The completion of this project will also represent the completion of the first two criteria set forth by Anders Sandberg and Nick Bostrom in their paper Whole Brain Emulation: A Roadmap [1], which would imply that we will be well on our way toward our first functional brain emulation. With the apparent eminence of, at least a simplistic version, of whole brain emulation, we must begin to consider some implications for the future. The goal of whole brain emulation is the eventual use of the technology to emulate a human mind. We must consider who the first person we choose to emulate will be. That means we must determine what traits will result in the best emulation, and, once these traits have been enumerated, what process should be used to find a person who possesses these traits and is willing to undergo the uploading procedure.

In the paper we will consider the criteria and selection process other groups have used in selecting their subject(s) and glean from them desirable attributes and procedures. Humanity is pushing the envelope, planning the creation of digital versions of ourselves. In this, the question of "who first?" is not trivial, quite the opposite, the answer will set the tone and the trajectory for future artificial superintelligence [2-4], and consequently the quality of our collective future [5]. We cannot take this decision lightly, we must attempt to set aside cultural bias/differences, and make a choice together, as a global community, taking our most breathtaking step forward.

## 2. Whole Brain Emulation

Whole brain emulation also referred to as "uploading" [6] or sometimes "downloading" is a process in which the structure of a brain is scanned in such a way that a software model could be created from that scan [1]. This software representation would be to such a level of sameness, that when the program is run, it will behave in an identical fashion to the original brain. The brain emulation will be a one to one emulation of the original. Emulation in this context, refers to a computer program producing the same behavior as another program, by copying its low level functions, it achieves the desired outward behavior [7-9] by copying that which causes that outward behavior. This differentiates emulation from simulation, in that simulation just endeavors to copy the outward behavior [10], a reactive process, whereas emulation is a proactive process [1].

In this paper, we will be operating under the assumption that the resulting mind emulation will achieve a level of detail such that it will be indistinguishable from the original brain, and will therefore inherit the traits of the individual to whom the original brain belonged. We are also assuming that the mind emulation is dynamic, and will therefore continue separately from that individual and be shaped by its own experiences.

## 3. Why it Matters Whom to Upload First

As stated above, whole brain emulation will result in a one to one emulation of the original brain, therefore that emulation will begin its existence as an exact replica of the original personality. It will possess their memories and experiences, up to the point of the procedure. It will behave with their personality traits, internal value system, everything about us that makes us an individual, this emulation will possess (have an exact operational copy) of the original person. This is why it is imperative to make a rigorous, thoughtful effort when selecting the first individual(s) to be emulated. The first person(s) emulated will set the tone of a wholly new experience for our global community, and will either magnify or mitigate the possible side effects of a substrate independent mind.

*Safety*

It is inevitable that whole brain emulation will lead to superintelligence. Once a sufficient approximation of our own intelligence is created, what is to stop that entity from surpassing its creator's capacity for problem solving, it will be "born" with perfect, infallible memory, and the ability to process information, that takes us hours, in a fraction of a second. It will, most likely, have access to the internet and all the information stored therein [11]. Therefore, it will inevitably become more intelligent than the brightest among us, more intelligent than we can comprehend at this juncture. Hopefully this new superintelligent being will be benevolent, grateful to those that gave it "life". However, it is easy to imagine the opposite occurring, and being deathless, it seems a high probability that at some point this entity will become malicious to us its creators [12].

The creation of a computer based approximation of human intelligence opens the door to some serious questions concerning our safety. It is not difficult for one to imagine a military application for such a device. A weapon, with the intelligence of a human, without the need for sleep, or rest of any sort. This application would change, forever, the way wars are fought. It would be utopian to think that we would remove the human element from our conflicts, simply pitting one nations technology against another's, may the best executed win! No, if history is any indication of future application, we can be assured that the incorporation of digital warriors in our conflicts will only make them bloodier, more clandestine, with further reaching effects on the civilians in close proximity to that violence.

*Personality*
Even though personality is a difficult concept to define on paper, we all have an innate understanding of the concept. It would behoove us to delineate criteria for higher level functions, to make sure that some negative attributes will not manifest in the human emulation. We will certainly want to eliminate anyone who deviates significantly from normal behavior patterns, especially if those deviations manifest in destructive and or negative interactions with his or her fellows. It would be desirable to find a mind donor who possesses characteristics that could be considered passive, an aversion to conflict, and a strong affinity for intrapersonal relationships.

*Values*
Let us define values as Schwartz and Bardi do in their article *Value Hierarchies Across Cultures: Taking a Similarities Perspective*. Values are "…desirable, transsituational goals, varying in importance, that serve as guiding principles in people's lives." [13]. In the social sciences they have found 10 motivationally distinct types of values which comprehensively encompass the core values of human society across cultures, philosophical and religious beliefs [13]. These are - Power: Social status and prestige, control or dominance over people and resources. Achievement: Personal success through demonstrating competence according to social standards. Hedonism: Pleasure and sensuous gratification for oneself. Stimulation: Excitement, novelty, and challenge in life. Self-direction: Independent thought and action choosing, creating, exploring. Universalism: Understanding, appreciation, tolerance, and protection for the welfare of all people and for nature. Benevolence: Preservation and enhancement of the welfare of people with whom one is in frequent personal contact. Tradition: Respect, commitment, and acceptance of the customs and ideas that traditional culture or religion provide the self. Conformity: Restraint of actions, inclinations, and impulses likely to upset or harm others and violate social expectations or norms. Security: Safety, harmony, and stability of society, of relationships, and of self [13].

Let us now introduce the concept of *subjective wellbeing*, research in this field looks to quantify how a person evaluates their life, both in the moment they are being evaluated and over an extended period of time [14]. It has been found that subjective wellbeing can be broken down into two main aspects, one that is cognitive and focuses on a person's sense of satisfaction with life in general, and one that is affective, and quantifies a person's feeling of happiness or sadness [15]. In the research done by Sagiv and Schwartz [16], they have determined that achievement, stimulation, and self-direction from the set of 10 values correlates to a positive sense of subjective wellbeing, while traditional values tend to correlate more negatively, and conformity and security correlated negatively with a person's sense of happiness (the affective aspect of subjective wellbeing) [15]. Interestingly, power, security, and conformity values were found to have very little, if any correlation to subjective wellbeing, suggesting that they are neither cause nor predictor of one's sense of positivity about life and person [16], [15].

With this in mind, we would certainly like to select an individual who is experiencing a high sense of subjective wellbeing, and has been experiencing that consistently over an extended period of time. Therefore we should look for individuals who are found to experience high levels of achievement, self-direction, and stimulation, while feeling less motivated by traditional values and power. For our unique purposes, it would be desirable to find an individual who is also motivated by the universalism and benevolence values, seeding the emulation with a sense of social justice and honesty.

## 4. Previous Work/Literature Review

Humanity will likely depend completely on the nature of the internal organization and the personality traits of the emulated brain. Consequently, it is vitally important to choose the brain wisely. The election would be similar to deciding what sort of god we would like to have for ourselves. In this section we survey history of humanity's "firsts" from leadership selection to space exploration, all the way to DNA donation.

**Leadership Selection**

So how do we choose the president? How about Supreme Court justices? Or perhaps the UN secretary general? In the cases of the president (at least the U.S. president), other than an age, residency and citizenship requirement, no specific requirements exist. This may work in the case of a large country with at least a few checks and balances on power, but we would not recommend a popular election to decide on the first person to upload. In the case of supreme court justices, according to Norman Dorsen in his article, "The Selection of U.S. Supreme Court Justices" [17], argues that the requirement is a balance between knowledge and understanding of constitutional law, "…the most valuable judicial qualities of competence, impartiality, empathy and wisdom." [17]. These qualities are more along the lines of what the first uploaded brain should possess.

In the case of selecting a secretary general for the UN, the criteria and qualifications for appointment are purposefully vague so as not to disqualify someone based on a structure that is unnecessarily rigid. Included in the requirements are the following: administrative and executive qualities, leadership qualities, and moral authority, political judgement, communication and representation skills, and …*"overall qualities which demonstrate to the world at large that personally the candidate 'embodies the principles and ideals of the Charter to which the Organization seeks to give effect'."* [18]. These are qualities that are undoubtedly important to the human race, but are not sufficient for accomplishing any of the goals that would make whole brain emulation a necessary risk. The human chosen for the first and maybe second uploads must also be a competent researcher in a field of interest. This would make the selection similar to that of Project Mercury, the process of choosing the first men to go into space.

*First Astronaut*

Project Mercury was the first American "man-in-space" program, which began in 1958 and was completed in 1963. The objectives of the project were to "orbit a manned spacecraft around the Earth, investigate man's ability to function in space, and to recover both man and spacecraft safely" [19]. The Mercury Project and the initial Apollo missions used similar criteria when selecting their first crew members.

Applicants to the Mercury, Gemini, and initial Apollo programs were all volunteers and required to have significant experience as pilots of high-performance aircraft. These machines were similar in nature to the spacecraft in that they "consist of complex propulsion, electrical, mechanical, and hydraulic systems." [20]. The applicants were also required to have backgrounds in science or engineering and, as the field was narrowed to a reasonable size, these applicants received training in digital computer theory, guidance and navigation, astronomy, and geology [20]. It was the goal of the program, and a requirement for each individual to "be able to perform all the piloting duties on the command module without the aid of the others." [20] interestingly in the Gemini and Apollo missions, the applicants were further narrowed to be no more than six feet to accommodate the limited size inside the space craft. An age restriction was implemented, applicants could be no more than 34 years, "to maximize the amount of time a man can actively participate in a flight crew ..." [20].

Those that met the above criteria and were selected to move forward in the process were subjected to "a one-year indoctrination program". Which contained approximately 570 hours of classroom activity [20] After the completion of the classroom activities that comprise the indoctrination program, each person was assigned a specialty, 75 percent of the activities after this assignment were devoted to that specialty. The remaining time was spent in simulations, flying aircraft, and survival activates [20].

As we determine who will be the best person(s) to be uploaded first, we can draw upon some of NASA's procedures. First of all, it must be stated, those considered for mental donation should be volunteers. In most of the first world, this would be a given, but we hope to have volunteers from across the globe and the willing informed consent of all who are considered must be guaranteed! We should also consider a "training" period for prospective donors as we narrow the list of candidates. NASA used this time to make sure their team was educated and confident in every aspect of the mission they were meant to accomplish. We could use a "training" period to get to know each potential donor better. If we are going to create a deathless version of an individual we should know that individual at least as well as each astronaut knew their space craft and mission. Furthermore, and possibly more important, a period of immersive learning could give each potential donner the time to really come to terms with the repercussions of the choice they are making, and hopefully cause withdrawal before the actual procedure becomes imminent and the research team must start the process of selection anew, or worse, the donor regrets their decision to be uploaded, but cannot undo the process or turn off the emulation due to ethical regulation.

NASA elected to instate an age limit on their candidates, so as to get a substantial career out of each, made necessary by the amount of training each astronaut requires to become "space ready". While we do not necessarily require any time from our mental donor after the procedure is complete, the question of an age restriction on volunteers is a prudent one to consider. It has been well documented, that after a point in our aging process our mental capacity tends to decline. The mental donor selected for uploading should be determined to be in their mental prime or in that stage of development when learning comes most easily. Therefore we propose that an age limit, on both ends of the spectrum, should be instated. For the lower end of the range, we consider age of majority, globally this ranges from 15 to 21 [21]. To allow for a more mature ability to make such a decision we suggest 21. For the upper limit we consider the age restrictions placed on pilots, 65 years (*Fair treatment of Experienced Pilot Act*) This seems reasonable considering the mental faculties required to operate a machine as complex as a plane.

**First DNA to be profiled**

Sequencing the human genome was a mammoth accomplishment for humanity. The Human Genome Project (HGP) was an internationally collaborative event, culminating in the complete mapping of the human gene sequence, or genome (see genome.gov). For this research to take place, subjects had to be selected to provide the raw material for the research teams to begin their work. For research of this kind, the benefits and risks to these subjects must be weighed, prior to any experimentation taking place. The benefits of the HGP are significant: a more thorough understanding of our basic biology, and, by default, our diseases. It was assumed that with the completion of the project, therapies for genetic diseases would be more effective, easier to create, and possible methods of prevention would result. Interestingly, the project would not directly benefit the volunteer donors, clinically or financially. Those that chose to divulge their genetic make-up would have to do so out of a sense of altruism [22].

The risks for these altruistic donors, may at first, seem minimal. However, the information resulting from the HGP was made available on a public database, and contains information with the potential to reveal disease susceptibility/propensity. If it were possible for the donor to determine which sequence was their own, this knowledge could cause them distress, depending on the results of their sequence. If this information were found out by the general populous or the individual's employer, insurer, etc., they could suffer discrimination and/or embarrassment. It was therefore, imperative to the project to keep their donors identity confidential, not only by standard procedure in maintaining confidentiality, but also, ensuring that a large enough group of donors was obtained to ensure each individual's anonymity [22]. It was also determined that staff of laboratories involved in the project would be prohibited from donating DNA, due to a concern that staff members would feel pressured to donate, along with a greater level of difficulty to maintain confidentiality [22].

Informed consent is a major part of acquiring human subjects, and for the HGP this issue proved to be less straight forward than for other human-subject based research projects. The nature of the research made anonymity and confidentiality impossible to absolutely guaranteed for large-scale DNA sequencing, along with the individual's inability to withdraw their genetic library from the public database, should they later wish to [22].

The HGP's work raises some interesting parallel's to the selection process for our first choice for emulation. While it was the HGP's major concern that the confidentiality of their donor's information could not be breached, confidentiality is an impossibility for our donor. This raises a number of concerns for that person, unwanted celebrity being among the more benign. More seriously, it is probable, there will be a significant number of individuals who are adamantly opposed to creating artificial life of any kind. These individuals may engage in peaceful demonstrations to get their message out, they may also lash out at the person after whom, the emulation was modeled.

The notion that donors to the Human Genome Project would be unable to remove their DNA library from the public data base, raises the questions "will it be possible for an individual to retract their uploaded mental information once the person emulation is functioning as a separate individual?" This is an issue we must deal with as fully as possible before beginning our selection process. We must be capable of making the options very clear to our volunteers. Unfortunately, it is impossible to disclose an exhaustive list of the risks to the individuals considering volunteering.

We cannot know what the results of our research will be used for in the future. With this in mind it is imperative that we obtain informed consent from each individual who puts their name forward to be considered. It may be wise to reference the HGP's informed consent form when creating one for the purpose of selecting candidates for Uploading.

The HGP raises a very interesting and valid concern involving members of their own staff's ability to donate to the project. We agree that all persons and family members of those involved with the research process, in any capacity, should be barred from being considered for mental donation. As mentioned in the above synopsis, a sense of pressure to volunteer would be inevitable. Not to mention, the pressure one would feel if working to emulate a friend, co-worker, or family member, it would certainly be impossible for the team to remain objective in such a circumstance.

**Xenotransplantation and the Declaration of Helsinki**
The medical field has had the most hands-on experience with research involving human subjects, and should be looked to for guidance regarding volunteer selection and the treatment of those volunteers. In fact, from the field of medicine we get the *Declaration of Helsinki,* a concise document that spells out the "ethical guidelines for physicians and other participant in medical research" [23]. The declaration begins with the following self-definition: *"The World Medical Association (WMA) has developed the Declaration of Helsinki as a statement of ethical principles for medical research involving human subjects, including research on identifiable human material and data." [24].*

Under this definition, our research into uploading clearly falls under the umbrella of research the Declaration of Helsinki is meant to influence. Arguably, any research done on the resulting person emulation would also fall under its guidance in so far as the emulation would be identifiable human data. It is important that we familiarize ourselves with its principles before beginning any research involving human subjects.

According to the FDA, xenotransplantation is "any procedure that involves the transplantation, implantation or infusion into a human recipient of either (a) live cells, tissues or organs from a nonhuman animal source, or (b) human body fluids, cells, tissues or organs that have had *ex vivo* contact with live nonhuman animal cells, tissues or organs" [25]. The controversial nature of xenotransplantation research, along with the many unknowns that accompany experimentation of this type on human subjects, required a rigorous screening process for the first trial recipients. First and foremost, the recipient must volunteer for and understand the risks associated with xenotransplantation.

Xenotransplantation is on the cutting edge of medical science and is therefore dealing with a lot of unknown risks and consequences for those who first undergo the procedure. The idea of possibly pursuing more than one path simultaneously, tailoring each to the unique needs of the individuals involved and/or different end goals, could be a beneficial practice when we begin to upload people. Having multiple teams, could combat tunnel vision, only going about the process in one way with one goal.

While "life"-long monitoring of the resulting emulation is a requirement of this type of research, we also need to be prepared to take care of the mind donor for the duration of their life. This person or persons will be in a situation like no one before them. Their privacy will be nonexistent, their deepest secrets, their most precious memories will be made available to another entity, and at the very least the staff that helped to develop that entity if not the public in general. We need to have the assets to assist this individual(s) to cope with their new reality. It will be a wholly unique experience and no one can be truly prepared for what it will be like to be that exposed. We must also keep in mind that this total transparency will not be limited to just the individual to whom the mind belongs, any person they have come in contact with will experience the same transparency in so far as their interactions with the donor. Therefore, we must consider extending informed consent and post procedural care to the donors close relations and family.

With the long list of side effects in store for anyone who decides to donate their mind for uploading, we must ask, what are the benefits? What would a person gain from being uploaded? That answer is simple, the first person emulated, would be the first person to achieve a form of immortality. A goal we have been chasing since we became aware of our mortality.

## 5. Selection Criteria

Taking into consideration the importance of choosing our first candidate(s) for uploading and referencing what we can learn from processes of selection used in other fields. Let us now take a look at what our selection criteria could look like.

**Diversity**

It is our hope that the first person emulation will be a global endeavor making a diverse pool of volunteers an inevitability, however it is more likely the task will be undertaken by a group of individuals that are much less diverse [26]. Whomever is conducting the research, they must not be tethered to just their countryman when casting their net for volunteers, this net should be cast globally! Hopefully we will have the means to upload multiple individuals in the first round of person emulations. Choosing a manageable number of individuals from a diverse assortment of ethnic and cultural backgrounds will have a number of benefits. First being, the benefits for human and person mind emulation interactions, a person is going to be much more likely to accept an artificial person if it "resembles" them. If the emulation speaks like them, knows their traditions, and cultural nuances, the technology will become quite relatable and less unnerving. Finally, creating multiple uploads of different individuals would be beneficial to the donors themselves. Rather, then being alone in this process, they would have a network of others who are dealing with the same unique situation.

**Disposition**

The mind donor's disposition will also play a role in the selection process. It would be best for the selected person(s) to have a predominantly positive view of themselves, humanity in general, and the future. They should be friendly and have shown a propensity to be charitable and helpful towards others. If these traits manifest in the emulation, we would create a "likeable" entity, one that would be more inclined to interact positively with the humans it comes in contact with. There is also the potential that the emulation will act in a more benevolent fashion towards its fellows. If the mind donor was compassionate towards those he or she came in contact with, it could only stand to increase the probability that the emulation would be more likely to act this way as well.

**Intelligence**

Let us consider Howard Gardener's Theory of Multiple intelligences, in which he proposed that intelligence is basically pluralistic. Gardener proposed seven types of intelligence, but was not married to these particular seven or even the notion that there shouldn't be a greater or lesser number. Intelligence, he theorized, "…is an ability to solve a problem or to fashion a product which is valued in one or more cultural settings" [27]. This definition allowed him to quantify abilities that were valued by a community, but could not be quantified by the intelligence tests of his day.

In selecting a mind donor for our first emulation, it would be wise to choose someone who possess a type of intelligence that we find desirable, and is difficult to learn. For an emulation, which at its heart is a computer program, spatial intelligence and the ability to problem solve should be relatively easy to "teach". There already exits programs to solve complicated mathematical equations and it is conceivable that we would be able to incorporate this type of program into the emulation. What about other desirable traits, that are more difficult to predict, quantify, and therefore code?  Gardner brings up the distinction between a person who is good at problem solving, and a person who is good at problem finding. Gardner made the point that coming up with a strong scientific theory is showing a propensity for problem finding, rather than solving and requires a wholly different skill set, a skill set that would be very difficult to code [27]. Therefore let us move this type skill, this type of intelligence to a higher priority when screening our pool of volunteers. Similarly we should prioritize a high level of interpersonal and intrapersonal intelligence, over high levels of, for example, spatial intelligence.

**Ethics**
It is important that all care is taken to treat all volunteers in an ethically sound fashion, this revolves around honesty, and transparency. Every aspect of the process (before, during, and after the procedure) must be laid out in a comprehensive informed consent form, in compliance with the Declaration of Helsinki (or similar relevant document of the time). Each volunteer must prove that they understand every aspect of the informed consent and must prove that they are of sound mind, capable of making an informed decision as well as volunteering of their own free will and under no coercion. All steps must be taken to have lifelong support set up for those that are selected for the procedure, as they will be in a position unlike any person has ever been. They will have, quite literally no privacy, no secrets from their life up to the point of uploading. The wellbeing and experience for the first persons uploaded will set the tone of public opinion and the ease at which future research will be performed.

**Potential suffering**
Unfortunately, there is a great deal of potential ~~for~~ suffering for the volunteers selected for uploading. It is very hard to imagine what the repercussions of making, literally every aspect of your inner life public. The decision to be uploaded will not just effect the one being uploaded, it will affect every person that individual has come in contact with. Especially those in closest proximity to the mind donor, in a very real way, they will be uploaded as well, and their privacy will also disappear. Embarrassment and guilt will not be in short supply, even for the most confident and open individuals.

There is also the possibility that despite our best efforts the resulting emulation turns out to be malevolent, and wreaks havoc on those individuals, with whom, it must deal. The technology embarked upon with these first uploads could set us on a path that makes all human labor obsolete, causing mass unemployment. These programs may one day organize and enslave or wipe out the human race. Although this outcome is not likely to occur immediately and with the great minds working on a solution to the containment for these emulations, we foresee this problem being manageable by the time uploading is possible!

## 6. Selection Process

The process we use for selecting our first uploadees should be a multistep experience. Initially we should cast a wide net, as wide a net as possible. It is important that the media we use to advertise for applicants is highly accessible, so as not to inadvertently screen out persons who can't afford the media we use. Therefore it would be wise to use a variety of media platforms, certainly social media (or its equivalent), but also the newspaper's equivalent, billboards, etc. whatever the project can afford. In this way, we will hopefully generate positive excitement for the project, and get a deep and diverse pool of applicants. In this first step, we will gather general data about each individual, such as age, ethnic background, general mental health, medical conditions that would affect their brain function, a short essay in which the applicant explains why they are interested in being uploaded. This application should contain a detailed description of what uploading is and what the potential risks are for any person who is uploaded (including nonphysical risks). Each applicant should be required to read this material.

Using the information gathered from the above applicants we should screen out those who do not meet age requirements, have suffered brain trauma, or do not meet requirements in some other, very obvious way. The remaining applicants will be asked to complete a number of tests, to determine their disposition and personality type. This second step in the process should be made a bit long and tedious, to screen out those who are not interested enough in the process to fill them out. Now, this shouldn't screen out those applicants who are unable to read, therefore a help network should be setup to allow those applicants who need it to call in and complete these surveys with the aid of a staff member (taking whatever means necessary to keep that staff administrator as nonbiased as possible.) This process should be used to find applicants that fit the personality, internal values, and intelligence profile desirable for our first upload.

The final stage of selection is the most involved. Using the information gathered in the two previous surveys, a manageable number of applicants should be selected to come live with each other and the research team. In this time, the applicants should be observed in their interactions with each other and researchers. In this time, applicants should be taught any skills they do not yet possess that would enhance the upload and a significant amount of time should be spent making sure that each applicant understands what it is they are doing and understands the consequences as we understand them at that time. This step in the selection process will allow the researchers to observe the applicants as they are naturally (not a dolled up version of themselves, that looks good on an application), it also allows the applicants time to fully come to terms with the repercussions of being the first person(s) uploaded, and if need be, remove themselves from consideration. Therefore this step should not be short, and should be as long as fiscally possible for the project. At the end of this time, of the candidates that remain, and have been deemed to fit all criteria for uploading, the first person(s) to be uploaded should be selected.

## 7. Whom Not to Upload

As important as it is to determine who would make a good candidate for uploading, we must also consider who would not. We should not upload any person who has suffered significant brain injury or deviates significantly from "normal" brain function. This would exclude persons who have suffered from stroke or are presently suffering from dementia or other degenerative brain disorders. Persons diagnosed with certain psychological disorders should also be excluded from consideration. This should include those diagnosed with antisocial personality disorder, narcissistic personality disorder, schizophrenia, bipolar disorder, borderline personality disorder, etc. Individuals convicted of any violent crimes should also be excluded.

No individuals from the research team or immediately related to the research team should be considered either. As discussed in the section on the human genome project, the inclusion of these individuals could result in feelings of obligation to apply for consideration, along with a compromised level of performance/judgement by the research staff members themselves if they are too closely related to/involved with the mind donor.

## 8. Conclusions

In the case of deciding which brain to upload first, there are several aspects to take into consideration. First to be taken into consideration is what we want to accomplish with this first brain emulation. Do we just want to see if it is possible, or is there a hierarchy of important tasks that we believe a faster brain is more suited for? What is at the top of the hierarchy? A successfully uploaded and emulated brain can have many copies, all thinking about the same problem and at much faster speeds with perfect recall of past thoughts and perfect records of past experiments. This would be a very large team of scientists working without any need for lunch breaks or naps. These copies can help with a myriad of problems, but which should be first?

The next thing to consider, is whether or not we will have control and accountability of the emulation. In the case that the emulation has unlimited power, we need to ask ourselves if the risks are worth the benefit of accomplishing our task. Since this is not (yet) an AGI, many scientists feel that we can understand and limit the power of a Whole Brain Emulation, giving it goals that the original human brain identified with and also valued. Nonetheless structures would need to be in place so that the emulation can be held accountable for its deliverables and safe for humanity. Having thought about accountability, it is also important to consider what kind of rights and autonomy will be given to the emulation and to the human it came from - if that human is still alive. It can be argued that ethical robotic engineering should not include emotions for robots, however, in the case of an emulated human brain it is difficult to imagine that the emotions of the human brain are not present. If the emulation chooses to stop working on the project because it no longer sees the value, is it ethical to terminate the emulation? If the premise is a contract between the original human and the researchers in charge of the emulation, is the emulation legally responsible to fulfill the contract that it had no part in signing? Answering these questions will guide the understanding of who should be uploaded first, and who should be uploaded next.

Given that whole brain emulation becomes a true possibility and the lack of any other controls in place, the most important accomplishment would be the development of safeguards for humanity. Along with the attributes of the men chosen as the first astronaut, the person chosen as the first upload will have to be well versed and experienced in AI research and understand the importance and the high risk impact of a singularity. If possible, it should also be noted that a preference holds for uploading a team of AI safety researchers simultaneously to mitigate the possibility of a rogue emulation behaving unexpectedly and poorly. This research should be done as soon as possible, so as to have as much of a natural barrier to progress in the form of computing power, once human brains are successfully uploaded.

One final consideration, is the issue of equality once safety has been achieved. Since a successfully emulated brain would be a way to immortality, the issue of accessibility should be thought through as well. This may be a major issue if the planet becomes unlivable for biological humans, and the only survivors are those with uploaded brains. As brain uploading moves from an issue of safety to simply an activity for the wealthy, and then a necessity for the masses, the protection of our physical environment may lose its purpose. At this point, brain uploading may become the only form of survival and it may not be available to everyone. Should it be?

We feel we have made a strong case for a thorough process, in determining who will make the best first candidate(s) for whole brain emulation. These individuals should first and foremost be volunteers, informed of the risks and benefits of the procedure to the best of our knowledge at the time. The person(s) chosen for the first emulation(s) should be of an appropriate age, free of physical trauma to the brain, and they should be without mental disorders. This person(s), should be found to be compassionate and able to empathize with others. They should already possess a strong ability to interact with and relate to their community, as well as an awareness of and ability to communicate how they are feeling internally. Depending on what the end use of the emulation will be, we should also add a propensity for skill sets that pertain to that end use. It is important to exclude persons too closely related to the research as well, to avoid unintended feelings of obligation.

There are still many unknowns when it comes to achieving a human emulation. Not least among these being that we don't know definitively how much of our *self* will be captured by a one to one structural replica of our brain. There is a notion of self that many of us believe exists apart from the physical structure that makes us who we are. It is exciting to think that the successful emulation of a person will bring us closer to understanding this dynamic. This uncertainty should not lessen the importance of whom we choose to upload first, for there are equal measures of possibility that our *self* will be uploaded with the more base brain functions, and that it will not. The consequence of not being thorough in selecting our first mind donor and ending up with an undesirable result, outweighs the toil required to engage in a rigorous selection process and not uploading the *self*.